\documentclass[aip,pop, reprint]{revtex4-1}
\usepackage{bm}
\usepackage{amsmath,amsfonts,amssymb}
\usepackage{blkarray}
\usepackage{color}
\usepackage{graphicx}

\newcommand{\vq}{\mathbf}

\newcommand{\B}{\textrm{B}}

\newcommand{\diff}{\mathop{}\!d}

\draft 

\begin{document}


\title{Plasma hydrodynamics from mean force kinetic theory} 

\author{Jarett LeVan}
\email[]{jarettl@umich.edu}
\affiliation{Applied Physics, University of Michigan, Ann Arbor, Michigan 48109, USA}
\affiliation{Nuclear Engineering \& Radiological Sciences, University of Michigan, Ann Arbor, Michigan 48109, USA}

\author{Scott D. Baalrud}
\email[]{baalrud@umich.edu}
\affiliation{Applied Physics, University of Michigan, Ann Arbor, Michigan 48109, USA}
\affiliation{Nuclear Engineering \& Radiological Sciences, University of Michigan, Ann Arbor, Michigan 48109, USA}

\date{\today}

\begin{abstract}
Mean force kinetic theory is used to evaluate the electrical conductivity, thermal conductivity, electrothermal coefficient, thermoelectric coefficient, and shear viscosity of a two-component (ion-electron) plasma. Results are compared with molecular dynamics simulations. 
These simulations are made possible by assuming a repulsive Coulomb force for all interactions. 
Good agreement is found for all coefficients up to a Coulomb coupling strength of $\Gamma \approx 20$. This is over 100-times larger than the coupling strength at which traditional theories break down. It is concluded that mean force kinetic theory provides a means to extend hydrodynamics to dense plasmas. 
\end{abstract}


\maketitle 

\section{Introduction}
Hydrodynamics of weakly magnetized plasmas requires the specification of six transport coefficients:~\cite{balescu,Chapman_Cowling,Ferziger_Kaper} electrical conductivity, thermal conductivity, thermoelectric coefficient, electrothermal coefficient, shear viscosity, and bulk viscosity. For classical weakly coupled plasmas, the Chapman-Enskog solution of the Boltzmann equation provides an excellent approximation for each of these.~\cite{Chapman_Cowling, Ferziger_Kaper, Damman_2025} In recent years, there has been considerable interest in extending plasma transport theory to dense plasmas in order to inform hydrodynamic simulations of inertial confinement fusion (ICF) implosions and a variety of astrophysical bodies.~\cite{Transport_workshop,LeePF1984,WhitePRL2020,DesjarlaisPRE2017,FrenchPRE2022,HainesPOP2024,SaumonPR2022} However, dense plasmas feature strongly correlated ions and degenerate electrons, both of which are difficult to describe in a first-principles theory. 

Mean force kinetic theory (MFKT), formalized in 2019,~\cite{Baalrud_2019} was developed to address the challenge of strong correlations. In MFKT, one solves a kinetic equation which resembles the Boltzmann equation, but where the potential of mean force replaces the Coulomb potential in the collision operator. The result is a kinetic equation that is free of divergences (unlike the traditional Landau and Boltzmann theories~\cite{Baalrud_2013}) and which  has matched results of first-principles molecular dynamics (MD) simulations to a far higher Coulomb coupling strength than traditional theories. In particular, the predictions of MFKT for shear viscosity,~\cite{Daligault_2014} thermal conductivity,~\cite{Scheiner_2019} and self-diffusion~\cite{Baalrud_2016} in a one-component plasma (OCP), as well as inter-diffusion in a two-component binary ionic mixture~\cite{BeznogovPRE2014,Shaffer_2017} were found to agree with MD results up to a coupling strength of $\Gamma \approx 20$, while the traditional theories fail at $\Gamma \approx 0.1$. Here, we are using the definition 
\begin{equation}
    \Gamma = \frac{e^2}{4\pi\epsilon_0 a k_\textrm{B} T}
\end{equation}
where $e$ is the electric charge, $\epsilon_0$ the permittivity of free space, $a = (3/4\pi n)^{1/3}$ is the average inter-particle spacing where $n$ is the total density, $k_\textrm{B}$ is Boltzmann's constant, and $T$ is the temperature.

In this work, we apply MFKT to the evaluation of transport coefficients in a two-component (ion-electron) plasma. Results for the electronic transport coefficients (electrical conductivity, thermal conductivity, thermoelectric and electrothermal coefficients) are compared with recent two-component MD calculations,~\cite{Damman_2025} and excellent agreement is found for $\Gamma \lesssim 20$. 
This is a similar range of agreement as has been observed previously for the OCP. 
The shear viscosity predicted from MFKT is compared with OCP data since no two-component calculations exist. Such a comparison is valid because shear viscosity is dominated by ions and has no diffusive contribution. Similar agreement with the MD is observed. No comparisons are made with the bulk viscosity because it is zero in kinetic theories and, to a good approximation, zero in classical atomic plasmas as well.~\cite{LeVan_BV}

To compare MFKT with the first-principles classical MD calculations, all interparticle forces are modeled as repulsive Coulomb. This choice is necessary because  attractive interactions in strongly coupled classical systems give rise to unphysical bound states.~\cite{Kuzmin_2002,Tiwari_2017} In the limit of weak coupling, this is not an issue and results are independent of the sign of the interaction, but in the strong coupling regime ($\Gamma \gtrsim 1$), the numerical values presented here are representative only of an ion-positron system. However, the importance of this work lies not in the numerical values of the coefficients, but in the demonstrated agreement between theory and MD for strongly coupled two-component systems. Specifically, the results show that MFKT captures strong-correlation effects on plasma transport up to $\Gamma \sim 20$, thereby addressing one of the two major challenges in developing a theory of dense plasmas. 

We note that MFKT has previously been applied to a classical attractive ion-electron system where the interaction was modified to include an artificial repulsive core at short-range to prevent bound states.~\cite{Shaffer_2019} Results of MFKT for the drift velocity relaxation rate in this system agreed well with MD.  
A quantum generalization of MFKT has also been developed for ion transport,~\cite{DaligaultPRL2016,StarrettPRE2013} electrical conductivity,~\cite{RightleyPRE2021,StarrettHEDP2017} and stopping power.~\cite{BabatiPRE2025} 
Similarly good agreement with density-functional theory molecular dynamics simulations~\cite{SjostromPRL2014,WittePOP2018,WhitePRB2018} have been observed for these cases.
A more comprehensive model, currently under development, will extend these works to consider all the hydrodynamics coefficients. 

The outline of this work is as follows: Section~\ref{sec:MFKT} reviews the mean force kinetic theory, Section III details the evaluation of the transport coefficients and compares the results with MD simulations, and Section IV summarizes the main results. 

\section{Mean force kinetic theory\label{sec:MFKT}}
Kinetic theories are typically formulated in terms of the velocity distribution function $f_s(\vq r, \vq v_s, t)$, defined such that $f_s(\vq r, \vq v_s, t) \diff \vq r \diff \vq v_s$ is the expected number of particles of species $s$ in $\diff \vq r$ about $\vq r$ with velocity $\diff \vq v_s$ about $\vq v_s$ at a time $t$. In MFKT, the evolution of $f_s$ is determined by~\cite{Baalrud_2019}
\begin{multline}
\label{eq:kineq}
    \biggl[\frac{\partial}{\partial t} + \vq v_s \cdot \frac{\partial}{\partial \vq r} + \frac{q_s}{m_s} \left(\vq E + \vq v_s \times \vq B\right) \cdot \frac{\partial}{\partial \vq v_s} \\ + \frac{1}{m_s} \overline{\vq F}^{(1)} \cdot \frac{\partial}{\partial \vq v_s}  \biggr] f_s(\vq r, \vq v_s, t) = \sum_{s'} \chi_{ss'}J(f_s f_{s'})
\end{multline}
where $\vq E$ is the electric field, $\vq B$ is the magnetic field, $q_s$ is the charge of species $s$, $m_s$ is the mass of species $s$, $\chi_{ss'}$ is the Enskog factor, and $J(f_s f_{s'})$ is the collision operator. The Enskog factor here is a generalization of that factor which appears in Enskog's theory of hard spheres to the case of continuous repulsive potentials.~\cite{Ferziger_Kaper,Baalrud_2015} It accounts for the fact that strongly repulsive particles have an effective excluded volume around them. For species of equal concentrations and the same interaction potential, the Enskog factor is species independent. See Ref.~\onlinecite{Baalrud_2015} for details. 

The quantity $\overline{\vq F}^{(1)}$ represents the mean force acting on a particle due to a statistically averaged background, given explicitly as
\begin{equation}
    \overline{\vq F}^{(1)}(\vq r) = \int \vq F_{12} \frac{n^{(2)}(\vq r, \vq r_{2})}{n^{(1)}(\vq r)} \diff \vq r_{2}
\end{equation}
where
\begin{equation}
    \vq F_{12} = \frac{e^2}{4\pi \epsilon_0} \frac{\vq r_1 - \vq r_2}{|\vq r_1 - \vq r_2|^3}
\end{equation} is the Coulomb force and $n^{(2)}$ and $n^{(1)}$ are the two particle and one particle local equilibrium density distribution functions. When the distribution functions are radially symmetric, which implies there is no virial pressure gradient,~\cite{Baalrud_2019} the integral vanishes. For this reason, we will set $\overline{\vq F}^{(1)} = 0$. Kinetic theories cannot yet account for virial contributions to transport, so if we were to allow the term to be non-zero, we would be modeling a system which has virial pressure gradients which never equilibrate.  

The collision operator in MFKT is
\begin{equation}
    J(f_s f_{s'}) = \iiint (\hat{f}_s \hat{f}_{s'} - f_s f_{s'}) u b \diff b \diff \epsilon \diff \vq v_{s'}
\end{equation}
where $\vq u = \vq v_{s'} - \vq v_s$ is the relative velocity between particles of species $s$ and $s'$, $b$ is the impact parameter, hatted quantities are evaluated at their post-collision velocities, and $\epsilon$ is the azimuthal angle which defines the direction of $ \hat{\vq g}$ relative to $\vq g$. In order to determine the post-collision state, a binary collision must be solved. In MFKT, collisions are mediated through the potential of mean force~\cite{Hansen_Mcdonald}
\begin{equation}
    w_{ss^\prime}^{(2)}(r) = - k_\textrm{B} T \ln [g_{ss^\prime}(r)]
\end{equation}
where $g_{ss'}(r)$ is the equilibrium radial distribution function. 
This is the average density profile particles of species $s^\prime$ surrounding a particle of species $s$ at the origin, normalized to the background density of species $s^\prime$. 
In this work, $g_{ss^\prime}(r)$ is evaluated using the hyper-netted chain (HNC) approximation.~\cite{HNC} Since all interactions are governed by the same repulsive potential, there is no need to distinguish between species when calculating $w_{ss^\prime}^{(2)}(r)$. Hence, the $g_{ss^\prime}(r)$ obtained here is the total radial distribution function. It is worth noting that $w_{ss^\prime}^{(2)}(r)$ reduces to the familiar Debye-Huckel potential in the limit of weak coupling. 

Hydrodynamic equations with explicit expressions for transport coefficients can be derived from kinetic equations by employing the Chapman-Enskog method.~\cite{Chapman_Cowling} Because the MFKT kinetic equation is structurally identical to Boltzmann's, one encounters little difficulty in applying the standard arguments. For a two-component quasi-neutral plasma, this procedure results in hydrodynamic equations of form~\cite{Ferziger_Kaper}
\begin{align}
    \frac{1}{\rho} \frac{d\rho}{dt} &= - \nabla \cdot \vq V, \\
    \rho \frac{d\vq V}{dt} &= - \nabla \cdot \vq P  + \vq j \times \vq B, \\
    \rho \frac{du_K}{dt} &= - \nabla \cdot \vq q - \vq P:\nabla \vq V + \vq j \cdot \vq E' ,
\end{align}
where $\rho$ is the total mass density, $\vq V$ is the center of mass velocity, $u_K = 3 nk_\B T / (2\rho)$ is the ideal internal energy per unit mass, $\vq E' = \vq E + \vq V \times \vq B$ is the electric field in the fluid reference frame, $\vq P$ is the pressure tensor, $\vq q$ is the heat flux, and $\vq j$ is the current density. These latter three quantities are determined by linear constitutive relations, which for a weakly magnetized plasma are given by~\cite{Ferziger_Kaper}
\begin{gather}
    \vq j = \sigma \left(\vq E' - \frac{1}{q_e n} \nabla p \right) - \varphi \nabla T, \\
    \vq q = -\lambda' \nabla T + \varphi T \left(\vq E' - \frac{1}{q_en}\nabla p \right) + \frac{5k_\B T}{2q_e} \vq j, \\
    \vq P = p \vq{I} - \eta \left[ \nabla \vq{V} + (\nabla \vq{V})^{\top} - \frac{2}{3}(\nabla \cdot \vq{V})\vq{I}\right]
\end{gather}
where $p=nk_\B T$ is the ideal scalar pressure, $\sigma$ is the electrical conductivity, $\lambda'$ is the partial thermal conductivity, $\varphi$ is the electrothermal coefficient, and $\eta$ is the shear viscosity. The system is coupled with Maxwell's equations, which in the non-relativistic limit are
\begin{subequations}
\begin{align}
\label{eq:maxa}
    \nabla \times \vq{E} &= - \frac{\partial \vq{B}}{\partial t}, \\
    \nabla \times \vq{B} &= \mu_0 \vq{j}, \\
    \nabla \cdot \vq{B} &= 0 .
\end{align}
\end{subequations}

The transport coefficients are written in terms of large integral equations, which one simplifies by expanding the unknown function in a series of Sonine polynomials of increasing order. Fortunately, the polynomial expansions converge quite fast and typically provide an excellent approximation by second order. The result is a set of expressions for transport coefficients in terms of bracket integrals
\begin{widetext}
    \begin{subequations}
    \begin{gather}
        [F,G]_{ss'}' = \frac{1}{2n_s n_{s'}} \iiiint f_{Ms} f_{Ms'} (G_s - \hat{G}_s)(F_s - \hat{F}_s) u b \diff b \diff \epsilon \diff \vq v_s \diff \vq v_{s'} \\ 
        [F,G]_{ss'}'' = \frac{1}{2n_s n_{s'}} \iiiint f_{Ms} f_{Ms'} (G_s - \hat{G}_s)(F_{s'} - \hat{F}_{s'}) u b \diff b \diff \epsilon \diff \vq v_s \diff \vq v_{s'}
    \end{gather}
\end{subequations}
\end{widetext}
where $F$ and $G$ are some function of the molecular velocity. In particular, defining the dimensionless peculiar velocity $\bm{\mathcal{C}} = (m/2k_\textrm{B}T)^{1/2} (\vq v - \vq V)$, the functions $F$ and $G$ are always of the form
\begin{equation}
    S_{3/2}^{(q)} (\mathcal{C}^2) \bm{\mathcal{C}} \quad \mathrm{or} \quad S_{5/2}^{(q)} (\mathcal{C}^2) (\bm{\mathcal{C}}\bm{\mathcal{C}} - \frac{1}{3} \mathcal{C}^2 \vq I)
\end{equation}
where $S_{p}^{q}(x)$ is a Sonine polynomial with index $p$ of order $q$. These forms of $F$ and $G$ allow the 8D bracket integrals to be reduced to linear combinations of 3D $\Omega$-integrals, defined as
\begin{equation}
    \Omega^{(l, r)}_{ss'} = \left( \frac{k_\textrm{B} T}{2\pi m_{ss'}}\right)^{1/2} \int_0^\infty \exp(-\tilde{u}^2) \tilde{u}^{2r+3} Q_{ss'}^{(l)} \diff \tilde{u}
\end{equation}
where $m_{ss'} = m_sm_{s'}/(m_s+m_{s'})$ is the reduced mass, $\tilde{u} = (m_{ss'}/ 2k_B T)^{1/2} u$ is the dimensionless relative velocity, and momentum cross section
\begin{equation}
    Q_{ss'}^{(l)} = 2 \pi \int_0^\infty \{1 - \cos^l[\theta_{ss'}(b,\tilde{u})]\} b \diff b
\end{equation}
where $\theta_{ss'}$ is the scattering angle. The scattering angle is determined from the binary collision problem, and can be written in form
\begin{equation}
    \theta_{ss'} = \pi - 2b \int_{r_0}^\infty \frac{\diff r/ r^2}{[1 - (b^2/r^2) -  w_{ss'}^{(2)}(r)/(k_\mathrm{B}T \tilde{u}^2)]^{1/2}},
\end{equation}
where $r_0$ is the distance of closest approach, determined by the largest root of the denominator. The exact linear relations between the bracket integrals and $\Omega$-integrals are tabulated in various textbooks.~\cite{Chapman_Cowling, Ferziger_Kaper} 

Evaluating the $\Omega$-integrals requires some care when using the potential of mean force. At each iteration of the inner-most integral, one must perform a numerical root-find to determine $r_0$. Moreover, since $\theta_{ss'}$ requires integrating up to the divergence at $r_0$, this root-find must be performed to an accuracy approaching machine precision. This is done here by performing a cubic spline interpolation of $w^{(2)}(r)$, then using Brent's root-find method~\cite{Brent_2013} and integrating via Gauss-Kronrod quadrature.~\cite{Davis_2007}

Before advancing to the evaluation of the transport coefficients, it is convenient to define the generalized Coulomb logarithms
\begin{equation}
    \Xi^{(l, r)} = \frac{1}{\chi}\sqrt{\frac{m_{ss'}}{2\pi k_\B T}} \left( \frac{8 \pi \epsilon_0k_\B T}{e^2}\right)^{2} \Omega_{ss'}^{(l,r)}.
\end{equation}
These functions are dimensionless, include the Enskog factor, and are species-independent. This is because the potential of mean force is species-independent (all particles interact through the same potential) if they all have the same charge, which is assumed here. This implies that the scattering angle is also independent of species. Consequently, $\Omega_{ss'}^{(l,r)} \propto m_{ss'}^{-1/2}$. Moreover, relative to the ion-electron mass ratio, all generalized Coulomb logs are of the same order, i.e., 
\begin{equation}
    \Xi^{(l,r)} \gg \frac{m_e}{m_i} \Xi^{(t,v)}
\end{equation}
for any $l, r, t, v$. This fact will be used when taking mass ratio expansions on expressions for the transport coefficients. All transport coefficients will be expressed in terms of $\Xi^{(l,r)}$. 

\section{Evaluation of the transport coefficients}

In mixtures, a direct comparison between kinetic theory and MD requires some care. There are generally two ways to define a heat flux - one which includes a contribution from the diffusion of enthalpy ($\vq q$) and one which does not ($\overline{\vq q}$). Of course, since they are different quantities, $\vq q$ and $\overline{\vq q}$ are satisfied by different linear constitutive relations. In particular, $\vq q$ satisfies~\cite{LeVan_2025, degroot_mazur}
\begin{equation}
     \vq{q} = - \tilde{\lambda} \nabla T +\tilde{\phi}\left[\vq{E} - \frac{m_eT}{q_e} \nabla \frac{\mu_e}{T} \right]
     \label{eq:heatflux}
\end{equation}
where $\tilde{\lambda}$ is the coefficient of thermal conductivity, $\tilde{\phi}$ is the thermoelectric coefficient, and $\mu_e$ is the specific electron chemical potential, while $\overline{\vq q}$ satisfies~\cite{degroot_mazur}
\begin{equation}
     \overline{\vq q} = - \overline \lambda \nabla T + \overline \phi \left[\vq E - \frac{m_e}{q_e} \nabla_T \mu_e\right],
\end{equation}
where $\nabla_T$ indicates a gradient at constant temperature. The transport coefficients defined between these two expressions are generally not equivalent. 

The heat flux definition also affects the current density equation due to the presence of the electrothermal coefficient. In particular, using definition $\vq q$ for the heat flux, the proper expression for the current density $\vq j$ is~\cite{LeVan_2025}
\begin{equation}
    \vq{j} = \tilde{\sigma}\left[\vq{E} - \frac{m_eT}{q_e} \nabla \frac{\mu_e}{T} \right] + \tilde{\varphi} \nabla T, \label{eq:current}
\end{equation}
where $\tilde{\sigma}$ is the coefficient of electrical conductivity and $\tilde{\varphi}$ is the electrothermal coefficient. 

Many popular MD codes (e.g. LAMMPS~\cite{LAMMPS}) define a heat flux that coincides with $\vq q$. The transport coefficients computed this way correspond to linear constitutive relations of the form of Eq.~(\ref{eq:heatflux}) and Eq.~(\ref{eq:current}).

In kinetic theory, the definition of heat flux often (but not always) coincides with $\vq q$. In particular, when it is written 
\begin{equation}
    \vq q = \sum_s \frac{1}{2} m_s \int (\vq v_s - \vq V)^2 (\vq v_s - \vq V) f_s \diff \vq v_s
\end{equation}
the definitions are consistent. However, the linear constitutive relations are typically written in the form~\cite{Ferziger_Kaper}
\begin{equation}
\label{eq:q_CE}
    \vq q = -\lambda' \nabla T + \varphi T \left(\vq E - \frac{1}{q_en}\nabla p \right) + \frac{5k_\B T}{2q_e} \vq j
\end{equation}
and
\begin{equation}
\label{eq:j_CE}
    \vq j = \sigma \left(\vq E - \frac{1}{q_e n} \nabla p \right) - \varphi \nabla T,
\end{equation}
where $p = nk_\B T$ is the pressure and $\lambda'$ is the partial coefficient of thermal conductivity. Kinetic theories rely on the ideal gas law. Hence, to make a comparison with the transport coefficients as defined in the Green-Kubo relations, the ideal gas limit of Eq.~(\ref{eq:heatflux}) and Eq.~(\ref{eq:current}) must be taken. This means we set
\begin{equation}
    \nabla \frac{\mu_e}{T} = \frac{k_\B}{m_e p} \nabla p - \frac{5k_\B}{2m_eT} \nabla T
\end{equation}
in Eqs.~(\ref{eq:heatflux}) and~(\ref{eq:current}). Comparing the resulting constitutive relations with those from kinetic theory, Eq.~(\ref{eq:q_CE}) and Eq.~(\ref{eq:j_CE}), yields the following relations:
\begin{subequations}
\label{eq:CE_MD_relations}
\begin{align}
    \tilde{\sigma} &= \sigma \\
    \label{eq:electhermmd}
    \tilde{\varphi} &= - \varphi - \frac{5k_\B}{2q_e }\sigma \\
    \tilde{\phi} &= T \varphi + \frac{5k_\B T}{2q_e} \sigma \\
    \label{eq:thermmd}
    \tilde{\lambda} &= \lambda' + \frac{5k_\B T}{q_e} \varphi + \frac{25 k_\B^2T}{4q_e} \sigma. 
\end{align}
\end{subequations}
The relation between the thermoelectric and electrothermal coefficient $\tilde{\phi} = - \tilde{\varphi} T$ is a consequence of time-reversal symmetry and is known as an Onsager relation.~\cite{degroot_mazur} There is no ambiguity in the viscosity coefficients since to linear order, viscous and thermal dissipation do not couple.~\cite{degroot_mazur}

In what follows, the coefficients on the right side of Eq.~(\ref{eq:CE_MD_relations}) correspond to what is computed from the Chapman-Enskog solution of MFKT, while the coefficients on the left correspond to what is computed from MD using the Green-Kubo relations. 
A comparison is made by putting the Chapman-Enskog results into Eq.~(\ref{eq:CE_MD_relations}). 

\subsection{Electrical conductivity}
In kinetic theory, the coefficient of electrical conductivity ($\sigma$) can be written as a sum of partial electrical conductivities ($\sigma_{ss'}$) as~\cite{Ferziger_Kaper}
\begin{equation}
    \sigma = \sum_{s,s'} \sigma_{ss'}
\end{equation}
where
\begin{equation}
    \sigma_{ss'} = \frac{n_s q_s\rho_{s'}}{p} \left[ \frac{q_{s'}}{m_{s'}} - \sum_{s''} \frac{n_{s''} q_{s''}}{\rho}\right] D_{ss'}.
\end{equation}
Here, $\rho$ is the total mass density, $\rho_s$ is the mass density of species $s$, and $D_{ss'}$ are the multi-component diffusion coefficients. For an ion-electron system with $n_i = n_e = n/2$, one has the relations \begin{equation}
    D_{ee} = - \frac{m_i}{m_e} D_{ei} = - \frac{m_i}{m_e} D_{ie} = \frac{m_i^2}{m_e^2} D_{ii}.
\end{equation}
Then, dropping all terms small by a mass ratio, the electrical conductivity can be written in simpler form
\begin{equation}
    \sigma = \frac{nq_e^2}{4\pi k_\B T}D_{ee}.
\end{equation}
The electron-electron diffusion coefficient $D_{ee}$ in the $n^{\mathrm{th}}$ approximation is given by
\begin{equation}
    D_{ee} =  \frac{1}{2n} d_{e,0}^{e(n)}
\end{equation}
where $d_{e,0}^{e(n)}$ is determined from the solution of
\begin{equation}
\label{eq:lin_d}
    \sum_{s'} \sum_{q=0}^{n-1} \Lambda_{ss'}^{pq} d_{s',q}^{e(n)} = \frac{8}{25k_B} \left( \delta_{se} - \frac{\rho_s}{\rho} \right) \delta_{p0}
\end{equation}
with $s = i,e$ and $p = 0, 1,...,n-1$. The coefficients $\Lambda_{ss'}^{pq}$ are linear combinations of bracket integrals
\begin{align}
    \nonumber \Lambda_{ss'}^{pq} = \frac{2m_{s}^{1/2} m_{s'}^{1/2}}{75 k_B^2 T} \biggl\{ &\delta_{ss'} \sum_{s''}  \left[ S_{3/2}^{(p)}(\mathcal{C}^2) \bm{\mathcal{C}},S_{3/2}^{(q)}(\mathcal{C}^2) \bm{\mathcal{C}} \right]_{ss''}' \\ &+ \left[ S_{3/2}^{(p)}(\mathcal{C}^2) \bm{\mathcal{C}},S_{3/2}^{(q)}(\mathcal{C}^2) \bm{\mathcal{C}} \right]_{ss'}''\biggr\},
\end{align}
which obey the following symmetry properties
\begin{equation}
\label{eq:symmetry}
    \Lambda_{ss'}^{pq} = \Lambda_{s's}^{qp} \quad \mathrm{,} \quad \sum_{s} \Lambda_{ss'}^{0q} = 0 \quad \mathrm{and} \quad  \sum_{s'} \Lambda_{ss'}^{p0}=0.
\end{equation}
The linear system Eq.~(\ref{eq:lin_d}) must be supplemented by the condition
\begin{equation}
    \sum_{s} (\rho_s/\rho) d_{s,0}^{e(n)} = 0.
\end{equation}
Here, we are using the convention that the $n^{\mathrm{th}}$ approximation to a coefficient refers to keeping the $n$ leading order non-zero terms one obtains after performing the Sonine polynomial expansion and evaluating the relevant integrals. This is distinguished from the Sonine polynomial order $q$, because the $q = 0$ term in the expansion can contribute to the transport coefficients. For example, the second approximation ($n=2$) for the diffusion coefficient uses only the $q = 0$ and $q=1$ Sonine polynomials.

The system in Eq.~(\ref{eq:lin_d}) can be solved directly via Cramer's rule. One finds $d_{e,0}^{e(n)}$ in terms of $\Lambda_{ss'}^{pq}$, which are expressible as bracket integrals, then writes the bracket integrals in terms of the generalized Coulomb logarithms. With $d_{e,0}^{e(n)}$, one can solve for $D_{ee}$, which allows determination of the electrical conductivity $\sigma$. 

It is well-established that transport coefficients are converged to within a few percent of their true value in the $n=2$ approximation,~\cite{kaneko_1960} so this is what will be done here. Results will be presented in terms of the dimensionless conductivity $\sigma^*$, defined by
\begin{equation}
    \sigma^* = \frac{\sigma}{\omega_{pe} \epsilon_0},
\end{equation}
where $\omega_{pe} = \sqrt{q_e^2 n_e/\epsilon_0 m_e}$ is the electron plasma frequency. For convenience, we also provide the result of the $n = 1$ approximation. The general algebraic solution to Eq.~(\ref{eq:lin_d}) is too large to reasonably write here. However, a tractable expression can be obtained by taking a first-order expansion in the mass ratio. This yields in the $n=1$ approximation
\begin{equation}
\label{eq:sigma1}
    \sigma^{*(1)} = \frac{\sqrt{3\pi}}{\Gamma^{3/2} \Xi^{(1, 1)}}.
\end{equation}
For $n=2$,
\begin{equation}
    \sigma^{*(2)}  = - \sqrt{\frac{3\pi}{2}} \frac{1}{2\Gamma^{3/2}} \frac{A_\sigma}{B_\sigma}
\end{equation}
where 
\begin{equation}
    A_\sigma = 25 \sqrt 2 \, \Xi^{(1, 1)} + 4 (-5 \sqrt 2 \, \Xi^{(1, 2)} + \sqrt 2 \, \Xi^{(1, 3)} + \Xi^{(2,2)}),
\end{equation}
and
\begin{equation}
    B_\sigma = 2 (\Xi^{(1, 2)})^2 - \Xi^{(1, 1)} (2 \Xi^{(1, 3)} + \sqrt 2 \, \Xi^{(2, 2)}).
\end{equation}
The problem of calculating electrical conductivity in an ion-electron plasma is thus reduced to the evaluation of four generalized Coulomb logarithms.

\begin{figure}
    \centering
    \includegraphics[width=\linewidth]{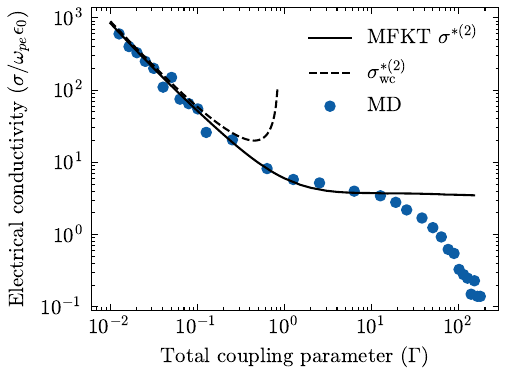}
    \caption{Mean force kinetic theory in the $n = 2$ approximation (solid line) and the weakly coupled plasma theory result (dashed line) for electrical conductivity compared with MD results from Ref.~\onlinecite{Damman_2025}. }
    \label{fig:conductivity}
\end{figure}

A comparison between MFKT, the traditional plasma theory, and MD is shown in Fig.~\ref{fig:conductivity}. For $\Gamma < 0.1$, the three are consistent. The traditional (weakly coupled) theory result is recovered within MFKT by replacing the generalized Coulomb logarithms with their weakly coupled limits~\cite{Baalrud_2014}
\begin{equation}
\label{eq:wc_cl}
    \Xi^{(l,r)} = l \Gamma(r) \ln \Lambda
\end{equation}
where $\Gamma(r)$ is the Gamma function and $ \ln \Lambda = \ln (\lambda_{D}/r_\textrm{L})$ is the traditional Coulomb logarithm, with $\lambda_{D}$ representing the total Debye length and $r_\textrm{L}=e^2/(4\pi\epsilon_ok_\B T)$ the Landau length. Hence, MFKT predicts that in the weakly coupled limit
\begin{equation}
    \sigma^{*(2)}_{\mathrm{wc}} = \frac{5.94}{\Gamma^{3/2} \ln \Lambda}.
\end{equation}
This can be written in more traditional form by defining the electron Coulomb collision time 
\begin{equation}
    \tau_e = \frac{3}{2\sqrt{2\pi}} \frac{\sqrt{m_e} (4 \pi \epsilon_0)^2 (k_B T)^{3/2}}{n_e q_e^4 \ln \Lambda}.
\end{equation}
Then we obtain the familiar weakly coupled textbook result~\cite{Ferziger_Kaper}
\begin{equation}
    \sigma_{\mathrm{wc}}^{(2)} = 1.93\frac{n_e q_e^2 \tau_e}{2m_e}.
\end{equation}
Of course this is not surprising because the potential of mean force reduces to the Coulomb potential in the limit of very weak coupling.   

For $0.1 < \Gamma < 20$, the electrical conductivity is shown to transition from  $\sigma \propto (\Gamma^{3/2} \ln \Lambda)^{-1}$ scaling to a plateau at 
\begin{equation}
    \sigma \approx 3.7 \epsilon_0 \omega_{pe} \propto \sqrt{n} \quad \mathrm{for} \quad 1 <\Gamma < 20.
\end{equation}
Physically, the plateau indicates that conductivity is independent of temperature for this range of coupling strengths. Traditional plasma theory does not predict this and begins to diverge from the MD at $\Gamma \approx 0.1$. However, MFKT shows excellent agreement with the MD data for this range of coupling strengths, meaning that if one fixes the density, MFKT applies to plasmas roughly 200 times cooler than the traditional theory. 

For $\Gamma > 20$, the electrical conductivity decreases and MFKT fails to capture this. At this large of coupling strengths, the system shows liquid-like behavior and the binary collision approximation breaks down.~\cite{Daligault_2006}Particles are trapped in local potential energy minima formed by neighboring particles and transport is dominated by the migration of particles across these minima.~\cite{Donk_2002}

\subsection{Electrothermal and thermoelectric coefficients}

The electrothermal coefficient ($\varphi$) can be written as a sum of partial electrothermal coefficients ($\varphi_{s}$) as
\begin{equation}
    \varphi = \sum_s \varphi_s
\end{equation}
where 
\begin{equation}
    \varphi_s = \frac{n_sq_s}{T} D_{T_s}
\end{equation}
and $D_{Ts}$ is the thermal diffusion coefficient. In an ion-electron system, one has the relation
\begin{equation}
    D_{Te} = - \frac{m_i}{m_e} D_{Ti}. 
\end{equation}
Hence, dropping $D_{Ti}$ since it is small by a mass ratio, we obtain
\begin{equation}
\label{eq:phidte}
    \varphi =  \frac{nq_e}{2T} D_{Te}.
\end{equation}
The electron thermal diffusion coefficient can be calculated from
\begin{equation}
\label{eq:dtea}
    D_{Te} =  \frac{\rho_i}{2\rho_e n} a_{i,0}^{(n)},
\end{equation}
where $a_{e,0}^{(n)}$ is obtained by solving the system
\begin{equation}
\label{eq:a_system}
    \sum_{s'} \sum_{q=0}^n \Lambda_{ss'}^{pq} a_{s',p}^{(n)} = \frac{2}{5k_B} \delta_{p1}
\end{equation}
for $s = i,e$ and $p = 0, ..., n$. This system is supplemented by the equation
\begin{equation}
\label{eq:constraint}
    \sum_s \frac{\rho_s}{\rho} a_{s,0}^{(n)} = 0.
\end{equation}
The process of solving for $\varphi$ follows similarly to what was outlined in the previous section. On defining the dimensionless electrothermal coefficient
\begin{equation}
     \varphi^* \equiv \frac{\varphi}{k_\B en_e/\omega_{pe}m_e},
\end{equation}
we find in the $n = 1$ approximation, after a first-order mass ratio expansion,
\begin{equation}
\label{eq:etherm1}
    \varphi^{*(1)}  =  \frac{5\sqrt{3 \pi} \left[5\Xi^{(1,1)} - 2 \Xi^{(1,2)}\right]}{\Gamma^{3/2} \left[4 (\Xi^{(1,2)})^2 - 2\Xi^{(1,1)} (2\Xi^{(1,3)} + \sqrt{2}\Xi^{(2,2)})\right]}.
\end{equation}
In the $n = 2$ approximation, even with a mass ratio expansion, the algebraic expression for $\varphi^{*(2)}$ is quite large. The reason is that Eq.~(\ref{eq:a_system}) is a larger system of equations than what was solved for the electrical conductivity. Moreover, the complexity of $\Lambda_{ss'}^{pq}$ grows as $p$ and $q$ grow. In particular, $\Lambda_{ss'}^{22}$ is a combination of nine distinct generalized Coulomb logarithms. The system is instead solved via Cramer's rule. 

To facilitate easy evaluation, we will write out the full system making use of Eq.~(\ref{eq:constraint}) and the symmetry properties of $\Lambda_{ss'}^{pq}$ given in Eq.~(\ref{eq:symmetry}),  and take a tedious term-by-term mass ratio expansion. The result is a reduced system $\vq M \vq x = \vq b$ where 
\begin{equation}
\label{eq:reduced_m}
    \vq M = \begin{bmatrix}
        \Lambda_{ii}^{00}\frac{m_i}{m_e} & \Lambda_{ie}^{01} & \Lambda_{ie}^{02} \\ \Lambda_{ie}^{01} \frac{m_i}{m_e} & \Lambda_{ee}^{11} & \Lambda_{ee}^{12} \\ \Lambda_{ie}^{02}\frac{m_i}{m_e} & \Lambda_{ee}^{12} & \Lambda_{ee}^{22}
    \end{bmatrix},
\end{equation}
\begin{equation}
    \vq x = \langle a_{i,0}^{(2)}, a_{e,1}^{(2)}, a_{e,2}^{(2)} \rangle ,
\end{equation}
and 
\begin{equation}
\label{eq:reduced_b}
    \vq b = \langle 0, \frac{2}{5k_B}, 0 \rangle.
\end{equation}
This was solved numerically for $a_{i,0}^{(2)}$, which with Eq.~(\ref{eq:phidte}) and Eq.~(\ref{eq:dtea}), leads to $\varphi^{*(2)}$. 

In the weakly coupled limit, application of Eq.~(\ref{eq:wc_cl}) yields
\begin{equation}
    \varphi_{\mathrm{wc}}^{(2)} = 0.69 \frac{k_B n_e q_e \tau_e}{m_e}.
\end{equation}
To compare the electrothermal coefficient with MD, the relation given in Eq.~(\ref{eq:electhermmd}) must be used. Here, we will use the ordering convention
\begin{equation}
    \tilde{\varphi}^{(n)} = - \varphi^{(n)} - \frac{5k_\B}{2q_e}\sigma^{(n)}.
\end{equation}
Note that $\varphi^{(n)}$ and $\sigma^{(n)}$ are evaluated to different orders in the Sonine polynomial expansion. 

In the MD, only the kinetic part of the electrothermal and thermoelectric coefficients were calculated. Generally, the heat flux contains contributions from the interactions between particles, leading to potential and virial contributions to transport coefficients. However, in a repulsive Coulomb mixture, the potential and virial contributions to these coefficients are ill-defined.~\cite{Damman_2025} Kinetic theories can only account for the kinetic contribution to transport, so one can still make a proper comparison. We note that the kinetic contribution to transport typically dominates up to $\Gamma \approx 20$.~\cite{Donko_1998,Daligault_2006,Daligault_2014, Scheiner_2019}

\begin{figure}
    \centering
    \includegraphics[width=\linewidth]{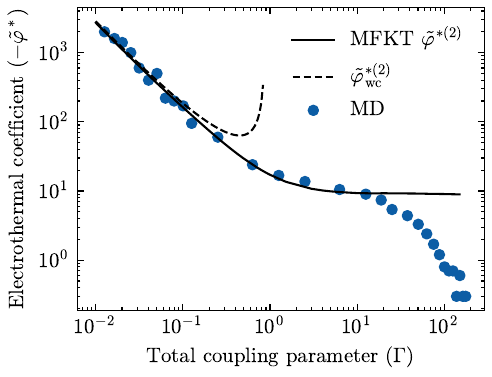}
    \caption{Mean force kinetic theory in the $n = 2$ approximation (solid line) and the weakly coupled plasma theory result (dashed line) for the electrothermal coefficient compared with MD results from Ref.~\onlinecite{Damman_2025}.}
    \label{fig:electherm}
\end{figure}

The comparison is shown in Fig.~\ref{fig:electherm}. The electrothermal coefficient is negative, so $-\tilde{\varphi}$ is plotted. Similar to the conductivity results, the traditional theory, MFKT, and MD all agree in the weakly coupled limit corresponding to $\Gamma \lesssim 0.1$. At moderate coupling strengths ($0.1 < \Gamma < 20)$, the MD shows a transition to a plateau at
\begin{equation}
    \tilde{\varphi} \approx -9.3 \frac{k_\B en_e}{\omega_{pe}m_e} \propto \sqrt{n_e} \quad \mathrm{for} \quad 1 <\Gamma < 20.
\end{equation}
This is captured by MFKT but not the traditional theory, and like electrical conductivity, is independent of temperature. For $\Gamma > 20$, the electrothermal coefficient decays and agreement with MFKT breaks down. 
It should be noted here, however, that potential and virial contributions would be expected to dominate in this regime in a physical system. 
These contributions are not included in the MD simulation analysis here because they diverge for a repulsively interacting system. 

\begin{figure}
    \centering
    \includegraphics[width=\linewidth]{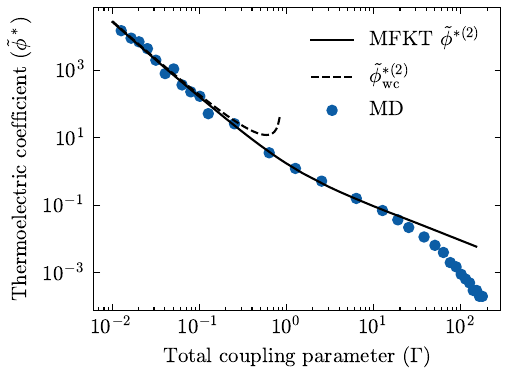}
    \caption{Mean force kinetic theory in the $n = 2$ (solid line) and the weakly coupled plasma theory result (dashed line) for the thermoelectric coefficient compared with MD results from Ref.~\onlinecite{Damman_2025}.}
    \label{fig:thermelec}
\end{figure}

The thermoelectric coefficient $\tilde{\phi}$ is related to the electrothermal coefficient via the Onsager relation
\begin{equation}
    \tilde{\phi} = - T \tilde{\varphi}.
\end{equation}
All results for $\tilde{\phi}$ are thus determined trivially in terms of $\tilde{\varphi}$. Of course, the thermoelectric coefficient shows similar agreement with MD, as it is directly related to the electrothermal coefficient. This is shown in Fig.~\ref{fig:thermelec}. Results are plotted in terms of the dimensionless thermoelectric coefficient
\begin{equation}
    \tilde{\phi}^* = \frac{\tilde{\phi}}{q_e \omega_{pe} / a_e}
\end{equation}
where $a_e = (3/4\pi n_e)^{1/3}$ is the average electron inter-particle spacing. 
Instead of a plateau, MFKT predicts a $1/\Gamma$ scaling at strong coupling, since the coefficient differs from the electrothermal coefficient by a temperature factor.  

\subsection{Thermal conductivity}
The partial coefficient of thermal conductivity is given in the $n^{\mathrm{th}}$ approximation by~\cite{Ferziger_Kaper}
\begin{equation}
    \lambda' =  \frac{5}{8}k_\B a_{e,1}^{(n)},
\end{equation}
where $a_{e,1}^{(n)}$ is obtained by solving the system in Eq.~(\ref{eq:a_system}). In writing this, terms small by an ion-electron mass ratio have been dropped. 

Defining the dimensionless partial thermal conductivity
\begin{equation}
\label{eq:lamdim}
    {\lambda'}^*  = \frac{{\lambda'}}{n_e \omega_{pe}k_\B a_e^2},
\end{equation}
in the $n = 1$ approximation
\begin{equation}
\label{eq:lam1}
    \lambda'^{*(1)} = - 
\frac{25}{(2)^{1/6}  \Gamma^{5/2}}\sqrt{\frac{\pi}{3}} \frac{A_{\lambda'}}{B_{\lambda'}}
\end{equation}
where
\begin{equation}
    A_{\lambda'} = \Xi^{(1, 1)}(-15 \Xi^{(1, 1)} + (4 + \sqrt{2})\Xi^{(2, 2)}),
\end{equation}
and 
\begin{align}
    B_{\lambda'} = (-2& (\Xi^{(1, 2})^2 + 2 \Xi^{(1, 1)} \Xi^{(1, 3)} + \sqrt{2}\Xi^{(1, 1)}\Xi^{(2, 2)}) \nonumber \\
    \times& (15\sqrt{2}\Xi^{(1, 1)} - 2(\Xi^{(2, 2)} + 2\sqrt{2} \Xi^{(2,2)})).
\end{align}
In the $n = 2$ approximation, the partial thermal conductivity must be evaluated numerically from the reduced system in Eqs.~(\ref{eq:reduced_m}-\ref{eq:reduced_b}). In the weakly coupled limit, one finds
\begin{equation}
    \lambda'^{(2)}_{\mathrm{wc}} = 2.07 \frac{n_e k_B^2 T \tau_e}{m_e}.
\end{equation}

\begin{figure}
    \centering
    \includegraphics[width=\linewidth]{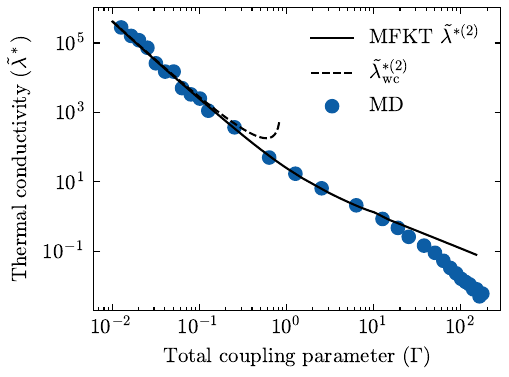}
    \caption{Mean force kinetic theory in the $n = 2$ (solid line) and the weakly coupled plasma theory result (dashed line) for the thermal conductivity compared with MD results from Ref.~\onlinecite{Damman_2025}.}
    \label{fig:thermal}
\end{figure}

To compare with MD, the full numerical solution is plugged into the previously obtained expression for $\tilde{\lambda}$ given in Eq.~(\ref{eq:thermmd}). The ordering convention
\begin{equation}
\label{eq:lambda_n}
    \tilde{\lambda}^{(n)} = \lambda^{'(n)} + \frac{5k_\B T}{q_e} \varphi^{(n)} + \frac{25 k_\B^2T}{4q_e} \sigma^{(n)}
\end{equation}
is used.
This comparison is shown in Fig.~\ref{fig:thermal}. Results are presented in the same dimensionless units as $\lambda'^*$ was written in Eq.~(\ref{eq:lamdim}). As with the electrothermal and thermoelectric coefficients, the MD data shown only represents the kinetic part of the coefficients since the potential and virial parts are ill-defined.

Similar to the previous coefficients, for $\Gamma \lesssim 0.1$, the MFKT, MD, and traditional plasma theory result all agree well. In the transition to strong coupling, the traditional weakly coupled plasma theory diverges, while MFKT shows good agreement up to $\Gamma \approx20$. From this, we conclude that MFKT can predict the electronic transport coefficients at coupling strengths two orders of magnitude larger than the traditional theory.   
It is interesting to note that the diffusive contributions to Eq.~(\ref{eq:lambda_n}) are dominant. 
That is, the term proportional to $\sigma^{(n)}$ is largest, and in fact gives an approximation to the total thermal conductivity that is generally accurate to within $\sim 50\%$ for all coupling strengths.
After this, the term proportional to $\varphi^{(n)}$ is most significant. 
The partial thermal conductivity is a relatively minor contribution to the total $<15\%$. 

\subsection{Shear viscosity}
The coefficient of shear viscosity is determined in the $n^{\mathrm{th}}$ approximation by
\begin{equation}
    \eta = \frac{1}{4}k_\B T b_{i,0}^{(n)} + \frac{1}{4}k_\B T b_{e,0}^{(n)}
\end{equation}
where coefficients $b_{s,0}^{(n)}$ are determined from the solution to 
\begin{equation}
    \sum_{s'} \sum_{q=0}^{n-1} H_{ss'}^{pq} b_{s',q}^{(n)} = \frac{1}{k_\B T} \delta_{p0}
\end{equation}
with $s = i,e$ and $p = 0,..., n-1$. The coefficients $H_{ss'}^{pq}$ are given by
\begin{equation}
\begin{split}
    &H_{ss'}^{pq} = \frac{1}{10k_B T} \\ &\times\biggl\{  \delta_{ss'} \sum_{s''}\left[ S_{5/2}^{(p)}(\mathcal{C}^2) (\bm{\mathcal{C}}\bm{\mathcal{C}} - \frac{1}{3} \mathcal{C}^2), S_{5/2}^{(q)}(\mathcal{C}^2) (\bm{\mathcal{C}}\bm{\mathcal{C}} - \frac{1}{3} \mathcal{C}^2)\right]_{ss''}'\\ &+ \left[ S_{5/2}^{(p)}(\mathcal{C}^2) (\bm{\mathcal{C}}\bm{\mathcal{C}} - \frac{1}{3} \mathcal{C}^2), S_{5/2}^{(q)}(\mathcal{C}^2) (\bm{\mathcal{C}}\bm{\mathcal{C}} - \frac{1}{3} \mathcal{C}^2)\right]_{ss'}'' \biggr\}.
\end{split}
\end{equation}
The $H_{ss'}^{pq}$ have the symmetry property
\begin{equation}
    H_{ss'}^{pq} = H_{s's}^{qp}.
\end{equation}
In the $n=1$ approximation, the system can be solved analytically. On defining the dimensionless viscosity
\begin{equation}
    \eta^{*} = \frac{\eta}{m_i n_i a_i^2 \omega_{pi}},
\end{equation}
one finds
\begin{equation}
\label{eq:eta1}
    \eta^{*(1)} = \sqrt{\frac{\pi}{3}} \frac{5(2)^{5/6}}{3\Gamma^{5/2}\Xi^{(2,2)}}.
\end{equation}
In the $n=2$ approximation, the system should be evaluated numerically using Cramer's rule. In matrix form, making use of the given symmetry, the system is $\vq M \vq x = \vq b$ where
\begin{equation}
    \vq M = \begin{bmatrix}
        H_{ii}^{00} & H_{ii}^{01} & H_{ie}^{00} & H_{ie}^{01} \\
        H_{ie}^{00} & H_{ei}^{01} & H_{ee}^{00} & H_{ee}^{01} \\
        H_{ii}^{01} & H_{ii}^{11} & H_{ei}^{01} & H_{ie}^{11} \\
        H_{ie}^{01} & H_{ie}^{11} & H_{ee}^{01} & H_{ee}^{11}
    \end{bmatrix},
\end{equation}
\begin{equation}
    \vq x = \bigl\langle b_{i,0}^{(2)}, b_{i,1}^{(2)}, b_{e,0}^{(2)}, b_{e,1}^{(2)} \bigr\rangle ,
\end{equation}
and
\begin{equation}
    \vq b = \biggl\langle \frac{1}{k_B T} ,\frac{1}{k_B T}, 0, 0 \biggr\rangle .
\end{equation}
In the weakly coupled limit, the numerical solution is
\begin{equation}
    \eta^{(2)}_{\mathrm{wc}} = 0.90 n_i k_B T \sqrt{\frac{m_i}{2m_e}}\tau_e.
\end{equation}

The MFKT prediction will be compared with MD results of shear viscosity in the one-component plasma because the authors are not aware of any ion-electron calculations. Since shear viscosity is associated with a momentum flux, it is dominated by ions due to their larger mass. Hence, the viscosity in the two systems is expected to be nearly identical if the density is taken to be just the ion density component when comparing with the one-component system.  

\begin{figure}
    \centering
    \includegraphics[width=\linewidth]{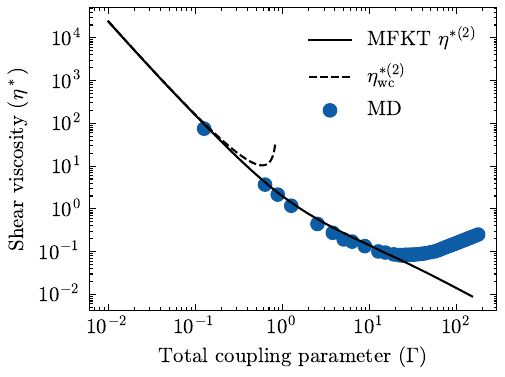}
    \caption{Mean force kinetic theory in the $n=2$ approximation (solid line) and weakly coupled plasma theory result (dashed line) for the shear viscosity compared to MD of the one-component plasma from Ref.~\onlinecite{Daligault_2014}}
    \label{fig:viscosity}
\end{figure}

The comparison between MFKT and MD is shown in Fig.~\ref{fig:viscosity}. Contrary to the previous coefficients, the shear viscosity from MD here includes the total, rather than only the kinetic part.~\cite{Daligault_2014} 
The minimum around $\Gamma =20$ is due to a transition from the kinetic components being most significant (for lower $\Gamma$ values) to the potential components (for higher $\Gamma$ values).~\cite{Daligault_2014} 
Good agreement with the MD data is observed up to this transition point, while the traditional theory diverges around $\Gamma \approx 0.1$. Notably, relative to the other coefficients, there appears to be a slight systematic over-prediction of the theory relative to the MD. Part of this many be due to an improper comparison. MFKT considers the impact of electron screening but the OCP model does not. However, even if one evaluates MFKT for a one-component system, a similar over-prediction is observed.~\cite{Daligault_2014} This may be an artifact of uncertainties in identifying a plateau in the cumulative time integral that one must do when evaluating Green-Kubo relations from MD, particularly since the shear stress autocorrelation function is slow to converge.

\subsection{Order comparison}
All transport coefficients in this work were evaluated in the $n=2$ approximation, meaning that the two leading-order terms in the Sonine polynomial expansion that yield non-zero contributions to the transport coefficients were used. Here, it is shown that in strong coupling ($\Gamma > 1$), the $n = 1$ approximation is well-converged for all coefficients. This means that the algebraic expressions given in Eqs.~(\ref{eq:sigma1}),~(\ref{eq:etherm1}),~(\ref{eq:lam1}), and (\ref{eq:eta1}) for the electrical conductivity, electrothermal coefficient, thermal conductivity, and shear viscosity respectively, are sufficient if $\Gamma \gtrsim 1$. 

\begin{figure}
    \centering
    \includegraphics[width=\linewidth]{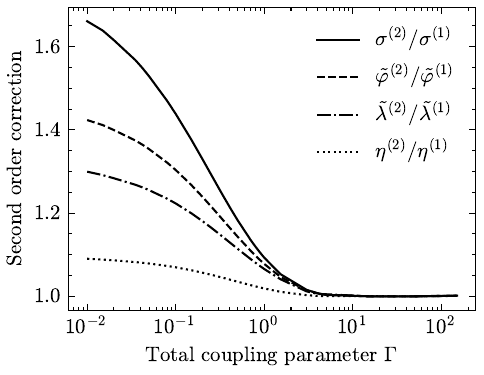}
    \caption{Ratio of the $n=2$ to $n=1$ approximation for the electrical conductivity (solid), electrothermal coefficient (dashed), thermal conductivity (dash-dotted), and shear viscosity (dotted) as computed from mean force kinetic theory. }
    \label{fig:ordercomp}
\end{figure}

A comparison between the first and second order results is shown in Fig.~\ref{fig:ordercomp}. For each coefficient, a ratio of the $n=2$ approximation to the $n=1$ is shown. For $\Gamma \geq 1$, the ratio is $\lesssim 1.1$ for all coefficients. As $\Gamma \rightarrow0$, each ratio can be evaluated using the weakly coupled limit of the generalized Coulomb logarithms from Eq.~(\ref{eq:wc_cl}). The results are
\begin{align}
\sigma_{\mathrm{wc}}^{(2)}/\sigma_{\mathrm{wc}}^{(1)}&=1.93, \quad   \quad \tilde{\varphi}_{\mathrm{wc}}^{(2)}/\tilde{\varphi}_{\mathrm{wc}}^{(1)} =1.53, \\
\tilde{\lambda}_{\mathrm{wc}}^{(2)}/\tilde{\lambda}_{\mathrm{wc}}^{(1)} &= 1.39, \quad \quad \eta_{\mathrm{wc}}^{(2)}/\eta_{\mathrm{wc}}^{(1)} =1.13.
\end{align}
Therefore, in MFKT's regime of validity for ($\Gamma \lesssim 20$), error in the $n=1$ approximation does not exceed $50\%$.

\section{Conclusions}
In this work, the electrical conductivity, thermal conductivity, electrothermal coefficient, thermoelectric coefficient, and shear viscosity were evaluated in a two-component ion-positron plasma from mean force kinetic theory. Results were compared against molecular dynamics and, for all coefficients, good agreement was obtained for a Coulomb coupling strength of $\Gamma \lesssim 20$. The traditional Boltzmann kinetic theory was shown to apply only in the limit $\Gamma \lesssim 0.1$. These results demonstrate that mean force kinetic theory can properly account for strong correlations in two-component plasmas. 
The model breaks down at $\Gamma \gtrsim 20$, where a transition to liquid-like behavior is expected to occur.\cite{Daligault_2006} 
The comparison with MD considered only a repulsively interacting (positron-ion) system because an attractively interacting system is not thermodynamically stable at strong coupling in a classical description, and because the sign of the charges is not expected to influence the results at weak coupling. 
Treatment of dense charge-neutral plasmas in the strongly correlated regime, such as warm dense matter, requires accounting for electron degeneracy, which is responsible for thermodynamic stability. 
Future work will explore applying a similar analysis to a quantum generalization of MFKT,~\cite{RightleyPRE2021,BabatiPRE2025} to treat dense plasmas. 


\begin{acknowledgments}
The authors thank Louis Jose and Lucas Babati for helpful conversations on this work. 
This work was supported in part by the NNSA Stewardship Science Academic Programs under DOE Cooperative Agreement DE-NA0004148, and the DOE NNSA Stockpile Stewardship Graduate Fellowship through cooperative agreement DE-NA0004185.
\end{acknowledgments}

\section*{Author declarations}
The authors have no conflicts to disclose.

\section*{Data Availability Statement}
Data sharing is not applicable to this article as no new data were created or analyzed in this study. 

\bibliography{aipsamp}

\end{document}